\documentclass[a4paper]{spie}
\usepackage[]{graphicx}
\usepackage{amsmath,amsfonts,amssymb}
\usepackage[colorlinks=true, allcolors=blue]{hyperref}

\title{Electromagnetic Imaging with Atomic Magnetometers: A Novel Approach to Security and Surveillance}

\author{Sarah Hussain, Luca Marmugi, Cameron Deans and Ferruccio Renzoni
\skiplinehalf
Department of Physics and Astronomy, University College London, Gower Street, London WC1E 6BT, United Kingdom}

\authorinfo{Send correspondence to F.~Renzoni: f.renzoni@ucl.ac.uk.\\ Pre-print version. Article reference: Proc. SPIE \textbf{9823}, Detection and Sensing of Mines, Explosive Objects, and Obscured Targets XXI, 98230Q (May 3, 2016); \href{http://dx.doi.org/10.1117/12.2222547}{DOI: 10.1117/12.2222547}.}

\begin{document} 
\maketitle

\begin{abstract}
We describe our research programme on the use of atomic magnetometers to detect conductive objects via electromagnetic induction. The extreme sensitivity of atomic magnetometers at low frequencies, up to seven orders of magnitude higher than a coil-based system, permits deep penetration through different media and barriers, and in various operative environments. This eliminates the limitations usually associated with electromagnetic detection.
\end{abstract}


\section{INTRODUCTION}
\label{sec:intro}
With the ever increasing access to global travel and movement of goods, the demand for a non-ionising, non-destructive and non-invasive device which is capable of detecting a range of materials is growing. Despite the various surveillance methods currently employed, many may present health hazards such as exposure to ionising radiation, or their effectiveness can be impeded by creative concealment of regulated or undesirable goods.  

We therefore propose a non-invasive, tunable detector capable of imaging conductive objects concealed by conductive enclosures such as cargo holds or thick shieldings. The device relies on a Radio-Frequency Optical Atomic Magnetometer (RF-OAM) operated in the Magnetic Induction Tomography (MIT) modality\cite{cam, oamsens}. MIT has already proved its usefulness in a number of industrial applications such as eddy current testing, but has since also demonstrated applications in medicine, for example in the detection of cerebral stroke \cite{stroke} and of heart arrythmias \cite{scirep}. 

\section{Magnetic Induction Tomography with Optical Atomic Magnetometers}
\begin{figure}[htbp]
    \includegraphics[width=0.25\textwidth]{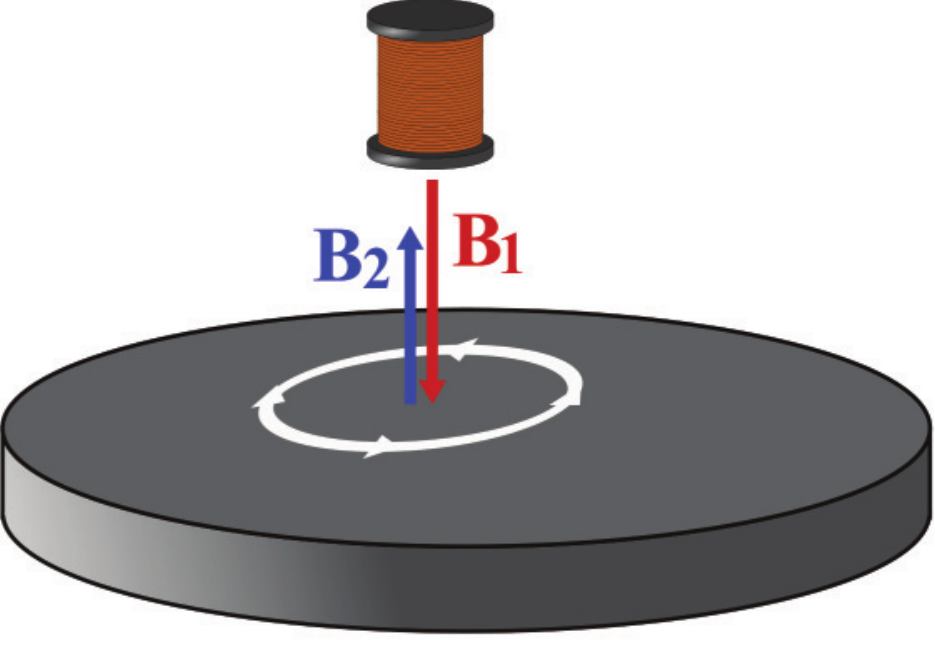}
    \centering
    \caption{MIT principle. $\mathbf{B_{1}}$, the primary magnetic field, is generated by a coil driven by an AC current source. It induces eddy currents (in white) into a conductive object, which generate a secondary field $\mathbf{B_{2}}$, whose amplitude and phase lag with respect to $\mathbf{B_{1}}$ depend on the dielectric properties of the supporting medium. A total magnetic field $\mathbf{B_{Tot}=B_{1}+B_{2}}$ is detected by a dedicated sensor.}
    \label{fig:mitter}
\end{figure}

MIT relies on the detection of a secondary magnetic field ($\mathbf{B_{2}}$) generated by eddy currents excited in a conductive target by a primary AC magnetic field ($\mathbf{B_{1}}$). The sketch in Fig.~\ref{fig:mitter} schematically represents this process. The secondary field contains information about the conductivity, permittivity and permeability of the object of interest \cite{mit}.

In conventional MIT systems, coils are used to produce eddy currents and also to pick-up the signal generated by $\mathbf{B_{2}}$. In our setup, based on magnetic induction imaging performed with an RF-OAM, the pick-up coil is replaced by an atomic magnetometer sensing unit \cite{cam}. A similar approach has recently been proposed by Wickenbrock \textit{et al}\cite{arnearxiv}.

The use of OAMs reduces or eliminates the impact of limitations and drawbacks of conventional sensors\cite{oamsens}, mainly in terms of sensitivity, available bandwidth and removal of capacitive coupling between induction and pick-up coils, which could alter the detection of the sample's properties. Alkali-metal vapor magnetometers have been shown to reach sensitivities which outperform SQUID magnetometers \cite{squidsens}. Moreover, the potential for miniaturisation for OAMs, thanks to micro-fabricated vapor cells, provides the potential for increased spatial resolution, reduced footprint and increased portability.

Other practical advantages include the operation of OAMs at room temperature and, in MIT modality with a phase-sensitive detection system, without the need for magnetic shielding. Finally, OAMs do not require calibration.

\section{Imaging Concealed Objects}
\begin{figure}[h]
\centering
    \includegraphics[width=0.7\textwidth]{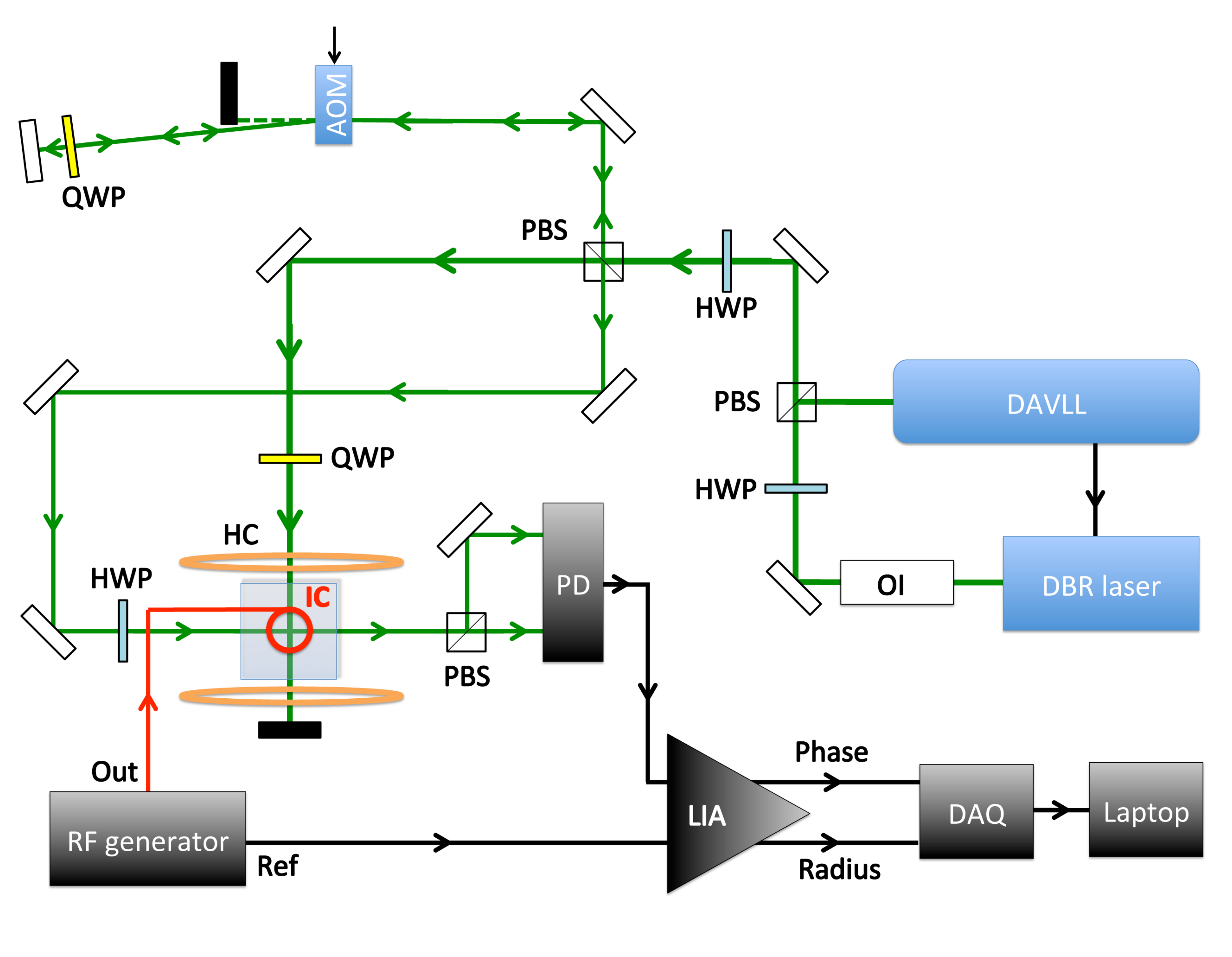}
    \caption{RF-OAM based MIT experimental setup. Key: OI is an optical isolator, PD is a balanced photodiode, AOM is an acousto-optical modulator, IC the induction coil generating the primary field and driving the OAM, HC is a Helmholtz coil pair, LIA is a lock-in amplifier and DAQ is the data acquisition device connected to a laptop. HWP and QWP refer to half and quarter waveplates, respectively. The sensing cell filled with a natural mixture of $^{87}$Rb and $^{85}$Rb is in the location of intersection of the pump and probe beams, between the Helmholtz coils. DAVLL is the dichroic atomic vapor laser lock used for laser's frequency stabilization.}
    \label{fig:setup}
\end{figure}

Our RF-OAM MIT setup has been extensively described elsewhere \cite{cam}; hence, we report here only the main details and a diagram of the setup (Fig.~\ref{fig:setup}). The RF-OAM is based on a rubidium vapor cell, where ${}^{87}$Rb is optically pumped and probed by laser radiation at 780 nm. 

The pump and probe laser beams intersect orthogonally at the center of the vapor cell. A smaller overlap potentially increases the spatial resolution of the MIT system since the only atomic spins involved in the operation are enclosed in that volume. In other words, the OAM would be sensitive only to local variations of $\mathbf{B_{Tot}}$. Nevertheless, this reduces the signal-to-noise ratio in given conditions. Therefore, a practical trade-off between spatial resolution and signal level must be found in view of desired applications.

The inductive coil (IC) sketched in Fig.~\ref{fig:setup} has two simultaneous functions. The first one is to coherently drive atomic precession in the RF-OAM. The second function is to induce eddy currents in the object of interest, as in conventional MIT configuration. The RF-OAM is tuned to resonance by adjusting the DC magnetic field generated by the Helmholtz coils pair, through suitable changes in the supplied current.

\begin{figure}[h]
\includegraphics[width=0.5\linewidth]{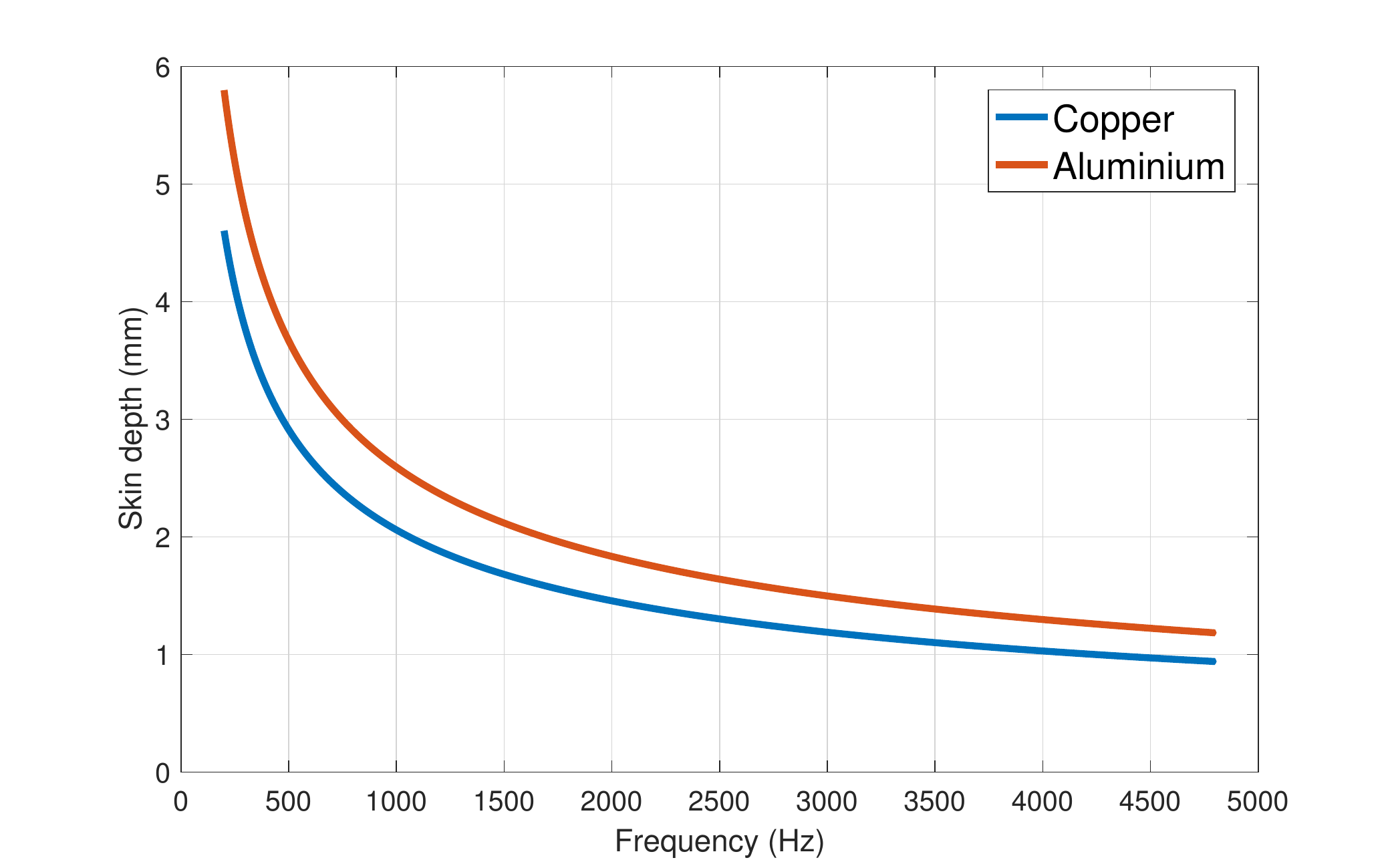}
\centering
\caption{Penetration depth through a conductive object as a function of frequency supplied to induction coil for copper ($\sigma$=59.8$ \times$10$^{6}$ S/m) and aluminium ($\sigma$=37.7$\times$10$^{6}$ S/m) samples.}
\label{skind}
\end{figure}

The eddy currents induced in the object of interest circulate mainly on the surface of the material, with the skin depth $\delta$ controlling their penetration. This is dependent on the inherent properties of the material itself and the frequency supplied to the induction coil. In the case of conductive and non-magnetic materials, it is given by \cite{sdepth}:

\begin{equation}
\delta=\sqrt{\dfrac{2}{\omega \mu_{0}\sigma}}~,\label{skinnyd}
\end{equation}

\noindent where $\sigma$ is the DC conductivity of the material, $\mu_{0}$ is the permeability of free space and $\omega$ the angular frequency of the AC field probing the object. The corresponding dependency for copper and aluminium is shown in Fig.~\ref{skind}.  

The choice of operation frequency depends on both the materials and the desired penetration (Eq.~\ref{skinnyd}). Thanks to the bandwidth and flexibility provided by the RF-OAM, the frequency can be chosen in order to image objects hidden by conductive shields. Once the induction coil frequency has been selected, the RF-OAM is tuned accordingly.

The signal from the probe beam at the photodiode is due to the total magnetic field acting upon the OAM unit, $\mathbf{B_{Tot}}$; that is to say it contains information about the driving field due to the induction coil and the signal due to eddy currents circulating in the target itself. A lock-in amplifier (LIA) is used to extract the information encoded the in OAM output, from which the magnetic field's properties and hence the dielectric characteristics of the object of interest are inferred. In particular, amplitude (radius) and retardation (phase) of the secondary field's contribution are measured.

\section{Imaging of Concealed Conductive Objects}
Here, the capability of our RF-OAM to detect concealed objects \cite{cam}, is investigated in detail. A copper square with sides 25 mm and thickness of 1 mm was shielded by a larger aluminium square with sides 37 mm and thickness also of 1 mm.

The test objects were placed on a Perspex mount positioned directly above the center of the atomic sensor. The mount, attached to a translation stage, was moved by micrometer screws. At each position of the object, $10^{3}$ data points obtained from the LIA output were averaged and recorded. After averaging, data were stored in a 2D array by means of a DAQ connected to a laptop computer. Data were then processed using a nearest neighbour filter with a radius r=2, and displayed in color-coded maps.

\begin{figure}[h]
\centering
\includegraphics[width=1\linewidth]{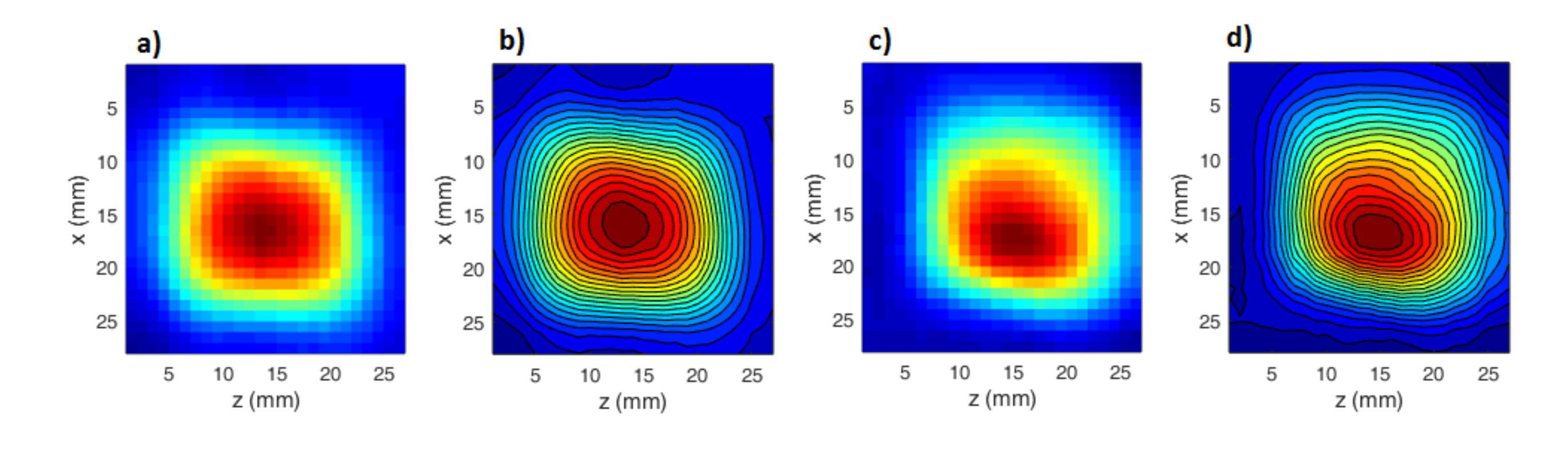}
\caption{MIT images of the unshielded copper square. \textbf{a)} and \textbf{b)}: radius image of the copper square and corresponding contour plot.  \textbf{c)} and \textbf{d)}: phase image and corresponding contour plot of the same sample. Images were taken with a 1 kHz driving field, which corresponds to $\delta_{Cu}=2.1$ mm in copper.}
\label{fig:noshield}
\end{figure}

The unshielded copper square images in Fig.~\ref{fig:noshield} accurately reproduce the size and shape of the Cu sample. The rounded shape of the image may be due in part to the nature of the circular pattern of eddy currents. The nearest neighbour filter also emphasizes this effect. 

The copper square was then shielded from the sensor using an aluminium square, with a thin insulating layer between them to prevent eddy currents being passed between the two materials, unlike the previous investigation \cite{cam}. The images were taken with a driving frequency of 500 Hz which corresponds to a skin depth of $\delta_{Al}=3.7$ mm in aluminium and $\delta_{Cu}=2.9$ mm in copper. The images were generated in a single measurement with no need for subtraction of the background. More importantly, no subtraction of the Al contribution was necessary, given the complete penetration of the AC fields through the concealing barrier.

\begin{figure}[h]
\centering
\includegraphics[width=1\linewidth]{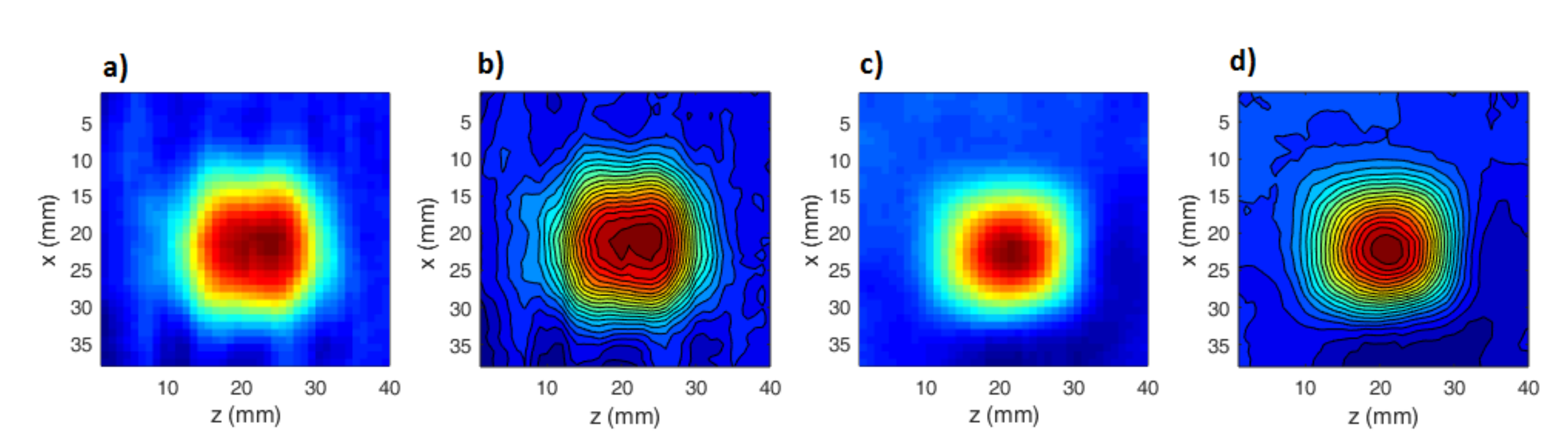}
\caption{MIT images of copper square shielded from induction coil with a larger aluminium square, with a non-conductive layer between the two metals. \textbf{a)} and \textbf{b)}: radius image and contour map of the shielded square. \textbf{c)} and \textbf{d)}: corresponding phase images.  Images were taken with a 500 Hz driving field, which corresponds to $\delta_{Al}=3.7$ mm and $\delta_{Cu}=2.9$ mm.}
\label{shieldedsquare}
\end{figure}

Results are shown in Fig.~\ref{shieldedsquare} for the radius and phase data respectively. By inspection, it is clear that the edges of the radius maps (panels a) and b)) are more noisy and hence less defined than those obtained in the phase maps (panels c) and d)). This deformation is particularly pronounced along the x axis. This has been attributed\cite{cam} to the presence of possible edge effects and mutual induction between the sample and the Al barrier.

Moreover, both maps appear smaller than those obtained in absence of the Al screen. This, in turn, can be explained by edge effects produced at the boundary of the two object, which may be particularly severe when the size of the samples are similar.

\section{Conclusion}
Our RF-OAM based MIT system has demonstrated the possibility of a real-time scanning device for the imaging and detection of conductive objects, with the ability to penetrate conductive shieldings. 

The sensitivity of our RF-OAM allows the imaging of conductive targets concealed by a conductive screen, without the need for magnetic shielding, or background subtraction. Furthermore, our system, thanks to the flexibility of the RF-OAM, can be easily tuned to image a wide variety of materials.

The current setup has a reduced complexity compared with the initial proof-of-principle demonstration \cite{wickenbrock2014}, reducing cost and size, thus improving everyday practical capabilities such as the potential for a hand-held scanner in an unshielded environment. This is ideal for applications in security and surveillance where a non-ionising, non-invasive easy-to-use scanner is required to detect a wide range of concealed objects. 

\section{Acknowledgements}
S.~H. is supported by DSTL - Defence and Security PhD - Sensing and Navigation using Quantum 2.0 technology. L.~M. is supported by Innovate UK as part of the AMMIT project. C.~D. is supported by the EPSRC Centre
for Doctoral Training in Delivering Quantum Technologies.

\end{document}